\documentclass[10pt,journal,letterpaper]{IEEEtran}%

\usepackage{amsfonts}
\usepackage{amssymb}
\usepackage{amsmath}
\usepackage{txfonts}
\usepackage[all,2cell]{xypic}

\usepackage{stmaryrd}

\IEEEpeerreviewmaketitle

\newtheorem{theorem}{Theorem}
\newtheorem{proposition}[theorem]{Proposition}
\newtheorem{corollary}[theorem]{Corollary}
\newtheorem{definition}[theorem]{Definition}
\newtheorem{example}[theorem]{Example}

\newcommand{\Z}{\mathbb{Z}}
\newcommand{\N}{\mathbb{N}}
\newcommand{\F}[2]{{\mathbb F}_{#1}^{\hspace{1pt} #2}}
\newcommand{\Cod}[1]{\mathcal{#1}}
\newcommand{\Conf}{\text{Conf}\left(A^{n}\right)}
\newcommand{\Equ}{\text{Equ}\left(A^{n}\right)}
\DeclareMathOperator{\id}{\text{\rm Id}}
\DeclareMathOperator{\aut}{\text{\rm Aut}}

\begin{document}

\title{On the Category of Group Codes}
\author{  \textbf{Rolando G\'omez Macedo} and \textbf{Felipe Zald\'ivar}%
\IEEEcompsocitemizethanks{\IEEEcompsocthanksitem R. G\'omez Macedo is with the Departamento de Matem\'aticas, Facultad de Ciencias, Universidad Nacional Aut\'onoma de M\'exico, 04510 M\'exico D. F., M\'exico. (e-mail: rolando@ciencias.unam.mx).}
\IEEEcompsocitemizethanks{\IEEEcompsocthanksitem F. Zald\'ivar is with the Departamento de Matem\'aticas,  Universidad  Aut\'onoma Metropolitana, 09340 M\'exico D. F., M\'exico. (e-mail: fz@xanum.uam.mx).}
\thanks{}}


\maketitle

\begin{abstract}
For the category of group codes, that generalizes the category of linear codes over a finite field, and with the generalized  notions of direct sums and indecomposable group codes, we prove that every MDS non trivial code, every perfect non trivial code, and every constant weight nondegenerate group code are indecomposable. We prove that every group code is a direct sum of indecomposable group codes, and using this result we obtain the automorphism groups of any group code in terms of its decomposition in indecomposable components. We conclude with the determination of the structure of decomposable cyclic group codes.
\end{abstract}


\section{\textbf{Introduction}}

Slepian \cite{Slepian} and Assmus \cite{Assmus} studied the category of linear codes over a finite field. In this work we extend this categorical formulation  to include codes defined over an arbitrary finite alphabet $A$. Here, $A^n$ is a metric space for the Hamming distance, and a code over the alphabet $A$ is a non empty subset ${\Cod{C}}\subseteq A^n$ with the induced metric. If $r\in \N$ and $c\in \Cod{C}$, the ball of center $c$ and radius $r$ is $B_r(c)=\{x\in A^n \mid d(x,c)\leq r\}$.  The key point is the definition of morphism. In \cite{Assmus} Assmus defines a morphism of linear codes, over a finite field, as a linear transformation  $\psi:\Cod{C}\rightarrow \Cod{D}$  such that $d(\psi(c_1),\psi(c_2))\leq d(c_1,c_2)$ for $c_1,c_2\in \Cod{C}$, where $d$ is the Hamming distance in the corresponding ${\mathbb F}_q^r$.
An immediate consequence is that $\psi:\Cod{C}\rightarrow \Cod{D}$ is an isomorphism if and only if it is an isometry. Several authors observed that this notion of isomorphism does not seem to take into account the number of errors that each of the involved codes corrects. 
To take this property into the definition Constantinescu and Heise \cite{Constantinescu} proposed the following: given linear codes $\Cod{C}\subseteq \F{q}{n}$, $\Cod{D}\subseteq \F{q}{m}$ over $\F{q}{}$, an isomorphism between  $\Cod{C}$ and $\Cod{D}$ is a linear isometry $\varphi:\Cod{C}\rightarrow \Cod{D}$ that is the restriction of a linear isometry $\varphi:\F{q}{n}\rightarrow \F{q}{m}$. By the MacWilliams Extension Theorem \cite{MacWilliams} and  \cite{MacWilliams_1}, see also \cite[Theorems 6.3 and 6.4.]{Wood}, the two previous definitions of isomorphism are equivalent in the category of linear codes.

Generalizing the above definitions to codes  over an arbitrary alphabet $A$, a {\it morphism} between codes $\Cod{C}\subseteq A^{n}$, $\Cod{D}\subseteq A^{m}$ is a map  $\varphi:A^{n} \rightarrow A^{m}$ such that $\varphi (\Cod{C})\subseteq \Cod{D}$ and  $d_{_{A^{m}}}(\varphi(x),\varphi(y))\leq d_{_{A^{n}}}(x,y)$ for all $x,y\in A^{n}$.
We say that $\varphi$ is an {\it isomorphism} if the map $\varphi:A^{n} \rightarrow A^{m}$ is bijective and its inverse $\psi$ is a morphism $\psi:{\Cod{D}}\rightarrow {\Cod{C}}$ of codes.

There is a natural notion of direct sum, since we have a bijection of $A^m\times A^m$  to  $A^{n+m}$, the latter has a Hamming distance given by $d=d_{A^n}+d_{A^m}$. If $\Cod{C}\subseteq A^{n}$ and $\Cod{D}\subseteq A^{m}$ are codes, its \textit{direct sum} is the code $\Cod{C}\oplus\Cod{D}=\{(x,y)\in A^{n+m}\mid x\in\Cod{C}, y\in\Cod{D} \}$. The following properties are immediate:
\begin{enumerate}
\item $\Cod{C}\oplus\Cod{D}$ has length $n+m$.
\item The minimum distance of the direct sum is $d_{_{A^{n+m}}}(\Cod{C}\oplus\Cod{D})=\min\left\{ d_{_{A^{n}}}(\Cod{C}), d_{_{A^{m}}}(\Cod{D}) \right\}$.
\item $\left|\Cod{C}\oplus\Cod{D} \right|= \left|\Cod{C} \right| \cdot \left|\Cod{D} \right| $.
\item $\Cod{C}\oplus\Cod{D}$ is isomorphic to $\Cod{D}\oplus\Cod{C}$.
\end{enumerate}
Let $\Cod{C},\Cod{C}^{\prime}\subseteq A^{n}$, $\Cod{D},\Cod{D}^{\prime}\subseteq A^{m}$ be codes and  $\varphi:\Cod{C}\rightarrow \Cod{C}^\prime$,  $\varphi:\Cod{D}\rightarrow \Cod{D}^\prime$ code morphisms. The  \textit{sum} $\varphi \oplus \psi:\Cod{C}\oplus \Cod{D}\rightarrow \Cod{C}^\prime\oplus \Cod{D}^\prime$ is the code morphism  given by $(\varphi \oplus \psi)(x,y)=(\varphi(x),\psi(y))$ for $(x,y)\in A^{n+m}$. If $\varphi$ and $\psi$ are isomorphisms, then $\varphi \oplus \psi$ is also an isomorphism.
Following Slepian \cite{Slepian} we say that a code $\Cod{C}\subseteq A^{n}$ is \textit{decomposable} if there are codes $\Cod{D}\subseteq A^{m}$ and $\Cod{E}\subseteq A^{l}$ such that $\Cod{C}$ is isomorphic to $\Cod{D}\oplus \Cod{E}$. If $\Cod{C}$ is not decomposable we say that  $\Cod{C}$ is an \textit{indecomposable code}.

In Section II we give examples of decomposable codes, in particular we will show that all non trivial MDS or perfect codes are indecomposable. We also give necessary and sufficient conditions for a code to be indecomposable.

When the alphabet is a finite group $G$ and the codes are subgroups of $G^n$, following Slepian \cite{Slepian} we call these codes {\it group codes}.
In Section  \ref{groucod} we study the category of group codes. The main result is that every group code has a decomposition as a direct sum of indecomposable group codes, unique up to isomorphism. Using this result we describe the automorphism group of a group code in terms of the automorphism groups of its indecomposable summands.

Throughout this paper we use the standard concepts: For a code ${\Cod{C}}\subseteq A^n$ over an alphabet $A$, its  length is $n$, its minimum distance $d({\Cod{C}})$ is the usual one, and its {\it dimension} is  $k=\log_q|{\Cod{C}}|$, where $q=|A|$ is the cardinality of the set $A$. In this situation we say that  ${\Cod{C}}$ is a $[n,k,d]_q$-code.
The usual Singleton bound holds: $k+d\leq n+1$. An MDS code (maximum distance separable code) is a code such that its parameters satisfy $k+d=n+1$.
 An integer $r\in \N$ is a \textit{correcting error radius} for $\Cod{C}$, if $B_{r}(c_1) \cap B_{r}(c_2)=\emptyset$ for all $c_1,c_2\in \Cod{C}$ with $c_1\neq c_2$. The largest correcting radius of a code  $\Cod{C}$ is $e=\left\lfloor\frac{d(\Cod{C})-1}{2}\right\rfloor$, an  it is called the \textit{correction capacity} of the code $\Cod{C}$.
 We say that $z\in A^{n}$ \textit{corrects the word} $c\in \Cod{C}$, if there exists a correcting error radius $r$ for $\Cod{C}$ such that $z\in B_{r}(c)$.
 A {\it perfect code} is a code $\Cod{C}\subseteq A^{n}$  such that any word of $A^{n}$ corrects  a word of $\Cod{C}$. A {\it trivial code} is a code isomorphic to some $A^n$.

\begin{proposition}\label{correcerroperf}
A perfect code $\Cod{C}$ with capacity correction  $e$ is perfect if and only if   $e$ is a correction radius for $\Cod{C}$ and for all $x\in A^{n}$ there exists $c\in \Cod{C}$ such that $x\in B_{e}(c)$.
\end{proposition}

\begin{proposition}\label{mdstrivial}\label{perfecttrivial}
{\rm (1)} If $\Cod{C}$ is a $\left[n,k,d\hspace{1pt}\right]_q$ MDS code over $A$, then $\Cod{C}$ is trivial if and only if  $d=1$.

\noindent{\rm (2)} A perfect code with correction  capacity  $e$ is trivial if and only if $e=0$.
\end{proposition}

If   $A$ is an alphabet  and $x_0\in A^{n}$ is a fixed element, the {\it weight} of $y\in A^{n}$ relative to  $x_0$ is $w_{x_0}(y)=d(y,x_0)$. If $x_0\in A^{n}$ and $0\leq r\leq n$ we denote by $R_{x_0}^{\hspace{1pt}r}=\{y\in A^{n}| w_{x_0}(y)=r \}$ the sphere of centre $x_0$ and radius $r$. A code $\Cod{C}\subseteq A^{n}$ is a \textit{constant weight code} if there exists  $x_0\in A^{n}$ such that $\Cod{C}\subseteq R_{x_0}^{\hspace{1pt}r}$.

We let $S_A$ denote the group of permutations of the set $A$. In particular, if $A=I_n=\{1,2,\ldots,n\}$, we set $S_{I_n}=S_n$. A function  $f:A^{n}\rightarrow A^{n}$ is a  \textit{configuration} of $A^{n}$ if there exist $f_1,\ldots,f_n\in S_A$ such that $f(a_1,\ldots,a_n)= (f_1(a),\ldots, f_n(a))$ for all $(a_1,\ldots,a_n)\in A^{n}$. An equivalence of $A^n$ is a map $\overline{\sigma}:A^{n}\rightarrow A^{n}$ induced by an element $\sigma\in S_n$ and given by  $\overline{\sigma}(a_1,\ldots,a_n)=(a_{\sigma(1)},\ldots,a_{\sigma(n)})$. Configurations and equivalences of $A^{n}$ are isometries of $A^{n}$. We denote by $\Conf$ the group of configurations of  $A^{n}$,  by $\Equ$ the group of equivalences of $A^{n}$ and by Iso$(A^{n})$ the group of isometries of  $A^{n}$.
 Markov Jr.\, see \cite[Theorem 14.2, pp. 300]{Hazewinkel} and Constantinescu and  Heise \cite{Constantinescu} have proven the following:

\begin{theorem}\label{teoestruismo}
If  $\varphi$ is an isometry of $A^{n}$, there exist $\overline{\sigma}\in \Equ$ and $f \in \Conf$ such that $\varphi=f \circ \overline{\sigma}$.
\end{theorem}

In general, if  $f=(f_1,\ldots,f_n)\in \Conf$ and $\sigma\in S_n$,  they induce a configuration $f_{\sigma}\in \Conf$ by means of $f_{\sigma}(x_1,\ldots,x_n)=(f_{\sigma(1)}(x_1),\ldots,f_{\sigma(n)}(x_n))$ for all  $(x_1,\ldots,x_n)\in A^{n}$. It follows that $\overline{\sigma}^{\hspace{1pt}-1} \circ f \circ\overline{\sigma}= f_{{\sigma}^{\hspace{1pt}-1}}$ and therefore
the group of configurations $\Conf$ is a normal subgroup of Iso$(A^{n})$. Moreover, since $\Conf \cap\Equ=\{\id_{A^{n}}\}$, then Iso$(A^{n})$ is a semidirect product:

\begin{corollary}\label{estrucgroautA^n}
 Iso$(A^{n})=\Conf\rtimes\Equ$.
\end{corollary}



\section{\textbf{Indecomposable codes over arbitrary alphabets}}\label{alfarb}

Given an arbitrary finite alphabet $A$, the {\it category of codes over $A$} has as objects the codes on the alphabet $A$. A {\it morphism} between two codes  $\Cod{C}\subseteq A^{n}$ and $\Cod{D}\subseteq A^{m}$ is a map  $\varphi:A^{n} \rightarrow A^{m}$ such that  $\varphi (\Cod{C})\subseteq \Cod{D}$ and moreover  $d_{_{\hspace{-1pt}A^{m}\hspace{-1pt}}}(\varphi(x),\varphi(y))\leq d_{_{\hspace{-1pt}A^{n}\hspace{-1pt}}}(x,y)$ for all $x,y\in A^{n}$. We denote this morphism by  $\varphi:\Cod{C}\rightarrow \Cod{D}$. Clearly, for any code $\Cod{C}\subseteq A^n$ the identity $\id:A^n\rightarrow A^n$ is a morphism $\id:\Cod{D}\rightarrow \Cod{D}$. The composition of two morphisms is also a morphism. An {\it isomorphism} of codes is a morphism $\varphi:\Cod{C}\rightarrow \Cod{D}$ such that the map $\varphi:A^n\rightarrow A^m$ is bijective and its inverse $\psi:A^m\rightarrow A^n$ is a morphism of codes $\psi:\Cod{D}\rightarrow \Cod{C}$. It follows that $n=m$ and that  $\varphi\circ \psi=\id_{\Cod{D}}$ and $\psi\circ \varphi=\id_{\Cod{C}}$. If there is an isomorphism between the codes $\Cod{D}$ and $\Cod{C}$ we will use the notation $\Cod{C}\simeq \Cod{D}$. Moreover, isomorphisms are restrictions of isometries of $A^{n}$ since for all $x,y\in A^n$,
$$d(x,y)=d(\psi(\varphi(x)),\psi(\varphi(y))\leq d(\varphi(x),\varphi(y))\leq d(x,y).$$
An {\it automorphism} is an isomorphism of a code onto itself.

\begin{example}
If  $\Cod{C}\subseteq A^{n}$ is a code and $1\leq m< n$, let $Y=\{i_{1},\ldots,i_m\}\subseteq I_n$, where $i_j\leq i_k$ for $j\leq k$. The function  $\pi_{_Y}: A^{n}\rightarrow A^{m}$ given by $\pi_{_Y}(x_1,\ldots,x_n)=(x_{i_1},\ldots,x_{i_m})$, determines the code $\pi_{_Y}(C)=\{\pi_{_Y}(c)\in A^{m}\mid c\in \Cod{C}\}\subseteq A^{m}$ and a morphism of codes $\pi_{_Y}:\Cod{C}\rightarrow \pi_{_Y}(C)$, called the $Y$-proyecci\'on of $\Cod{C}$.
\end{example}

\begin{example}
For codes $\Cod{C}\subseteq A^{m}$, $\Cod{D}\subseteq A^{l}$, and $b\in \Cod{D}$, a fixed element, the map $i_{b}:A^m\rightarrow A^m\oplus A^l$ given by $i_{b}(x)=(x,b)$ for $x\in A^{m}$, defines a morphism $i_{b}:\Cod{C}\rightarrow \Cod{C}\oplus \Cod{D}$. Observe that for all  $x,y\in A^{m}$ we have that $d(x,y)=d\left(i_{b}\left(x\right),i_{b}\left(y\right)\right)$. Similarly, if  $a\in\Cod{C}$ we have the morphism $i_{a}:\Cod{D}\rightarrow \Cod{C}\oplus \Cod{D}$ given by  $i_{a}(y)=(a,y)$ for $y\in A^{l}$.
\end{example}

\begin{example}
For any  alphabet $A$:
\begin{enumerate}
\item Every non empty subset of  $A$ is an indecomposable code.
\item If $\Cod{C}= A^{n}$, then $\Cod{C}=\bigoplus\limits_{i=1}^{n}\Cod{C}_i$, where $\Cod{C}_i=A$ for  each $i$, and $A$ is indecomposable.
\end{enumerate}
\end{example}

\begin{proposition}\label{MDSines}
Every non trivial MDS code is indecomposable.
\end{proposition}

\begin{IEEEproof}
Assume there exist an $\left[n,k,d\hspace{1pt}\right]_q$ nontrivial MDS code $\Cod{C}$ over $A$  and  codes  $\Cod{D}\subseteq A^{m}$, $\Cod{E}\subseteq A^{l}$ such that $\Cod{C}\simeq\Cod{D}\oplus \Cod{E}$. Without loss of generality we may assume that
 $d(\Cod{D})\leq d(\Cod{E})$. Since
 $\left| \Cod{D}\oplus \Cod{E} \right|=\left| \Cod{D}\right| \left|\Cod{E} \right|$, if $\Cod{D}$ and $\Cod{E}$ have parameters  $\left[m,k_1,d(\Cod{D})\hspace{1pt}\right]_q$ and  $\left[l,k_2,d(\Cod{E})\hspace{1pt}\right]_q$, then \linebreak$k= log_q\left(\left| \Cod{D}\oplus \Cod{E} \right|\right)=k_1+k_2$. Moreover, the parameters of $\Cod{D}$ y $\Cod{E}$  satisfy  $k_1+d(\Cod{D})\leq m+1 $ y $k_2+d(\Cod{E})\leq l+1$, respectively. Adding and using that $d(\Cod{C})=min\{d(\Cod{D}),d(\Cod{E})\}=d(\Cod{D})$, it follows that: \vspace{-3pt}
\[
k+d(\Cod{C}) + d(\Cod{E})=(k_1+k_1)+d(\Cod{D})+d(\Cod{E})\leq m+l+2=n+2. \vspace{-3pt}
\]
Finally, since $\Cod{C}$ is an MDS code, $d(\Cod{C})=d(\Cod{D})\leq d(\Cod{E}) \leq 1 $, by Proposition \ref{mdstrivial},  $\Cod{C}$ would be a trivial code, a contradiction.
\end{IEEEproof}

\begin{proposition}\label{perfines}
Every non trivial perfect  code is indecomposable.
\end{proposition}

\begin{IEEEproof}
Assume there exist an  $\Cod{C}\subseteq A^{n}$ a perfect non trivial code and $\Cod{D}\subseteq A^{m}$, $\Cod{E}\subseteq A^{l}$ codes such that $\Cod{C}\simeq\Cod{D}\oplus \Cod{E}$. Assume that these codes have error capacities
 $e_{_\Cod{C}}$, $e_{_\Cod{D}}$ and $e_{_\Cod{E}}$, respectively. Without loss of generality we may assume that
 $d(\Cod{D})\leq d(\Cod{E})$. Then, $e_{_\Cod{D}}=\left\lfloor\frac{d(\Cod{D})-1}{2} \right\rfloor  \leq \left\lfloor\frac{d(\Cod{E})-1}{2} \right\rfloor = e_{_\Cod{E}}$. In particular $e_{_\Cod{D}}$ is a correcting error radius for $\Cod{E}$. Moreover, since $d(\Cod{C})=min\{d(\Cod{D}),d(\Cod{E})\}$, then $e_{_\Cod{C}}=e_{_\Cod{D}}$. Let
 $a\in \Cod{D}$ and $x\in A^{l}$. Since $\Cod{D}\oplus \Cod{E}$ is a perfect code, there exists $(b,c)\in \Cod{D}\oplus \Cod{E}$ such that $d_{_{A^{n}}}((a,x),(b,c))\leq e_{_\Cod{C}}$. Therefore\vspace{-1mm}
{\fontsize{9.8}{12}
\[
d_{A^{l}}(x,c)\leq d_{A^{m}}(a,x)+d_{A^{l}}(b,c)=d_{A^{n}}((a,x),(b,c))\leq e_{_\Cod{C}}=e_{_\Cod{D}}\vspace{-1mm}
\]}
By Proposition \ref{correcerroperf}, $e_{_\Cod{D}}=e_{_\Cod{E}}$. Now consider $(\beta,\gamma)\in \Cod{D}\oplus \Cod{E}$  and let $y\in A^{m}$, $z\in A^{l}$ such that $d_{A^{m}}(\beta, y)=e_{_\Cod{C}}$ and $d_{A^{l}}(\gamma, z)=e_{_\Cod{C}}$. Since $\Cod{D}\oplus \Cod{E}$ is a perfect code, there exists $(\beta^{\prime},\gamma^{\prime})\in \Cod{D}\oplus \Cod{E}$ such that \vspace{-1mm}
\[
d_{A^{m}}(\beta^{\prime},y)+d_{A^{l}}(\gamma^{\prime},z)=d_{A^{n}}((\beta^{\prime},\gamma^{\prime}),(y,z))\leq e_{_\Cod{C}}=e_{_\Cod{D}}=e_{_\Cod{E}}.\vspace{-1mm}
\]
Therefore $d_{A^{m}}(\beta^{\prime},y)\leq e_{_\Cod{D}}$ and  $d_{A^{l}}(\gamma^{\prime},z) \leq e_{_\Cod{E}}$. Hence,  $y\in B_{e_{_\Cod{D}}}(\beta)\cap B_{e_{_\Cod{D}}}(\beta^{\prime})$ and $z\in B_{e_{_\Cod{E}}}(\gamma)\cap B_{e_{_\Cod{E}}}(\gamma^{\prime})$. It follows that   $\beta=\beta^{\prime}$, $\gamma=\gamma^{\prime}$ and
 \vspace{-2mm}
\[
2e_{_\Cod{C}}=e_{_\Cod{D}}+e_{_\Cod{E}}=d_{A^{m}}(\beta,x)+d_{A^{l}}(\gamma,y)\leq e_{_\Cod{C}}. \vspace{-2mm}
\]
Which, by Proposition \ref{perfecttrivial} is a contradiction since $\Cod{C}$ is a non trivial perfect code.
\end{IEEEproof}

The following is  a useful criterion:

\begin{proposition}\label{testparvesiseesci}
A code  $\Cod{C}\subseteq A^{n}$ is decomposable if and only if there exist $J,K\varsubsetneq I_n$ such that $J\cup K=I_n$, $J\cap K=\emptyset$ and  $\left| \Cod{C} \right|=\left|\pi_{_J}(\Cod{C})\right| \left| \pi_{_K}(\Cod{C})\right|$.
\end{proposition}

\begin{IEEEproof}
Assume that $\Cod{C}\subseteq A^{n}$ is a decomposable code. Then, there exist codes $\Cod{D}\subseteq A^{m}$, $\Cod{E}\subseteq A^{l}$, and  $\varphi: \Cod{D}\oplus \Cod{E}\rightarrow \Cod{C}$ an isomorphism such that $\varphi=f\circ \overline{\sigma}$, where   $f\in \Conf$ y $\overline{\sigma}\in \Equ$. If $b\in \Cod{E}$ is a fixed element, consider the inclusion $i_b:\Cod{D}   \rightarrow \Cod{D}\oplus \Cod{E}$. If $J=\sigma(I_m)$ and $K=\sigma(I_n-I_m)$, a straightforward computation shows that  $\pi_{_{J}} \circ\varphi\circ i_{b}:\Cod{D} \rightarrow \pi_{_{J}}\left(\Cod{C}\right) $ is an isomorphism. Similarly, if $a\in \Cod{D}$ is a fixed element, for the inclusion $i_{a}:\Cod{E}\rightarrow \Cod{D}\oplus \Cod{E}$, the composition $\pi_{_{K}} \circ\varphi\circ i_{a}:\Cod{E} \rightarrow \pi_{_{K}}\left(\Cod{C}\right) $ is an isomorphism. Hence, \vspace{-7pt}
\[
\left(\pi_{_{J}} \circ\varphi\circ i_{b}\right) \oplus\left(\pi_{_{K}} \circ\varphi\circ i_{a}\right):\Cod{D}\oplus\Cod{E}\rightarrow \pi_{_{J}}\left(\Cod{C}\right)\oplus \pi_{_{K}}\left(\Cod{C}\right)\vspace{-7pt}
\]
is an isomorphism and thus $\Cod{C}\simeq \pi_{_{J}}\left(\Cod{C}\right)\oplus \pi_{_{K}}\left(\Cod{C}\right)$.

Conversely, assume that $\Cod{C}\subseteq A^{n}$ is a code and that there exist   $J,K\subseteq I_n$ such that $\left| \Cod{C} \right|=\left|\pi_{_J}(\Cod{C})\right| \left| \pi_{_K}(\Cod{C})\right|$ and satisfy that $J\cap K=\emptyset$ y $J\cup K=I_n$. The last two conditions imply that  the map $\varphi:A^{n}\rightarrow A^{n}$  given by $\varphi(x)=\left(\pi_{_J}(x), \pi_{_K}(x)\right)$ is an isometry of $A^{n}$ such that $\varphi(\Cod{C})\subseteq\pi_{_J}(\Cod{C}) \oplus \pi_{_K}(\Cod{C})$. Since $\left| \Cod{C} \right|=\left|\pi_{_J}(\Cod{C})\right| \left| \pi_{_K}(\Cod{C})\right|=\left|\pi_{_J}(\Cod{C})\oplus \pi_{_K}(\Cod{C})\right|$, it follows that $\varphi\left(\Cod{C}\right)=\pi_{_J}(\Cod{C})\oplus \pi_{_K}(\Cod{C})$. Thus, $\varphi:\Cod{C}\rightarrow \pi_{_J}(\Cod{C})\oplus \pi_{_K}(\Cod{C})$ is an isomorphism.
\end{IEEEproof}

\begin{example}
If the alphabet $A=\Z/4$ is the ring of integers modulo $4$.  For the code $\Cod{C}\subseteq \left(\Z/(4)\right)^{3}$ given by   $\Cod{C}=\left\{(\overline{0},\overline{0},\overline{0}), (\overline{2},\overline{0},\overline{0}), (\overline{1},\overline{2},\overline{1}), (\overline{3},\overline{2},\overline{1}), (\overline{2},\overline{0},\overline{2}),
(\overline{0},\overline{0},\overline{2}),  (\overline{3},\overline{2},\overline{3}), \right. \\  \> \left.  (\overline{1},\overline{2},\overline{3})\right\}.$ Since
\begin{center}
\begin{tabular}{c @{ = } c @{ = } c}
  $\left|\pi_1(\Cod{C})\right| \cdot \left|\pi_{\{2,3\}}(\Cod{C})\right|$ &  $4\cdot 4$ & $16$ \vspace{2pt}\\
  $\left|\pi_2(\Cod{C})\right| \cdot \left|\pi_{\{1,3\}}(\Cod{C})\right|$ &  $2\cdot 8$ & $16$ \vspace{2pt}\\
  $\left|\pi_3(\Cod{C})\right| \cdot \left|\pi_{\{1,2\}}(\Cod{C})\right|$ &  $4\cdot 4$ & $16$ \vspace{3pt}\\
\end{tabular}
\end{center}
by Proposition \ref{testparvesiseesci}, $\Cod{C}$ is indecomposable.
\end{example}

\begin{proposition}\label{lemcanisomor}
Let $\Cod{C},\Cod{C}^{\prime}\subseteq A^{m}$ and $\Cod{D},\Cod{D}^{\prime}\subseteq A^{l}$. Assume that $\varphi:\Cod{C}\oplus\Cod{D}\rightarrow \Cod{C}^{\prime}\oplus\Cod{D}^{\prime}$ is an isomorphism and let  $f\in \text{Conf}(A^{m+l})$, $\overline{\sigma}\in \text{Equ}(A^{m+l})$ such that $\varphi=f\circ \overline{\sigma}$. If $\sigma(I_m)=I_m$, then $\Cod{C}$ is isomorphic to $\Cod{C}^{\prime}$, and $\Cod{D}$ is isomorphic to $\Cod{D}^{\prime}$.
\end{proposition}

\begin{IEEEproof}
Let $b\in\Cod{D}$ and consider the inclusion $i_{b}:\Cod{C}\rightarrow \Cod{C}\oplus\Cod{D}$. Since $\sigma\left(I_m\right)=I_m$, it follows that $\pi_{_{I_{m}}} \circ\varphi\circ i_{b}:\Cod{C} \rightarrow \Cod{C}^{\prime} $ is an isomorphism. From $\sigma\left(I_m\right)=I_m$ it follows that $\sigma\left(I_{m+l}-I_m\right)=I_{m+l}-I_m$, and if $a\in\Cod{C}$ is a fixed element and $i_{a}:\Cod{D} \rightarrow \Cod{C}\oplus\Cod{D}$ is the corresponding inclusion, then $\pi_{_{I_{m+l}-I_m}} \circ\varphi\circ i_{a}:\Cod{D} \rightarrow \Cod{D}^{\prime}$ is an isomorphism.
\end{IEEEproof}

\begin{definition}
A  code $\Cod{C}\subseteq A^{n+1}$ is {\it degenerated} if there exist $x\in A $ and $\Cod{D}\subseteq A^{n}$ such that
$ \Cod{C}\simeq\{x\}\oplus \Cod{D}$. Otherwise we say that $\Cod{C}$ is a \textit{non degenerated code}.
\end{definition}

\begin{corollary}
A code $\Cod{C}\subseteq A^{n}$ is degenerated if and only if there exists $i\in I_n$ such that $\left|\pi_i(\Cod{C})\right|=1$.
\end{corollary}

\begin{corollary}
If  $\Cod{C}\subseteq A^{n}$ is a non degenerated code of cardinality $p$, with $p$ a prime integer, then $\Cod{C}$ is indecomposable.
\end{corollary}


\section{Group codes}\label{groucod}

In this section we assume that the alphabet is a finite group $G$.
A {\it group code} is a subgroup $\Cod{C}\subseteq G^{n}$. If $\Cod{C}\subseteq G^{n}$ and $\Cod{D}\subseteq G^{m}$ are group codes, a {\it morphism of group codes} $\varphi:\Cod{C}\rightarrow \Cod{D}$ is morphism of codes that is also a homomorphism of groups. In particular, an {\it isomorphism of group codes} is an isomorphism of groups which is an isometry.

\begin{example}
If $I_n=\{1,2,\ldots,n\}$, for all  $\emptyset \neq Y\varsubsetneq I_n$ and any group code $\Cod{C}\subseteq G^{n}$, the projection \linebreak$\pi_{_{Y}}:\Cod{C}\rightarrow \pi\left(C\right)$ is a morphism of group codes. Given group codes $\Cod{D}\subseteq G^{m}$ and $\Cod{E}\subseteq G^{l}$, for the identity $\overline{e}_{l}\in G^{l}$, the inclusion $i_{\overline{e}_{l}}:\Cod{D}\rightarrow \Cod{D}\oplus\Cod{E}$ is a morphism of group codes.
\end{example}

Clearly, every $\overline{\sigma}\in \text{Equ}\left(G^{n}\right)$ is an automorphism of group codes. Hence, if $\varphi$ is an automorphism of the group code $G^{n}$, by Theorem \ref{teoestruismo}, there exist $\left(f_1,\ldots,f_n\right)\in \text{Conf}\left(G^{n}\right)$ and $\overline{\sigma}\in  \text{Equ}\left(G^{n}\right)$ such that $\varphi\circ \overline{\sigma}^{\vspace{1pt}-1}=\left(f_1,\ldots,f_n\right)$. Therefore, $f_i$ is an automorphism of the group code $G$.  From Corollary \ref{estrucgroautA^n} we obtain:

\begin{proposition}\label{autogroups}
Let $G$ be a finite group and $\Cod{C}\subseteq G^{n}$ be a group code. Denote by $\aut_{GC}(C)$ the automorphism group of the group code $\Cod{C}$, and by $\aut(\Cod{C})$ the automorphism group of the group $\Cod{C}$. Then,
 $\aut_{GC}(G^{n})= \left(\aut(G) \right)^{n}\rtimes  \text{Equ}\left(G^{n}\right)$.
\end{proposition}

Our main result shows that in the category of group codes, the indecomposable codes determine the group codes and the morphisms between them:

\begin{theorem}\label{estrucdecoddege}
Every group code $\Cod{C}$ is isomorphic to a direct sum of indecomposable codes, that is
$\Cod{C}\simeq \Cod{D}_1\oplus\cdots \oplus\Cod{D}_r$, where each $\Cod{D}_i$ is an indecomposable group code. This decomposition is unique up to permutation of the factors and isomorphisms, that is, if  we also have that $\Cod{C}\simeq \Cod{D}^{\prime}_1\oplus\cdots \oplus\Cod{D}^{\prime}_s$, where each $\Cod{D}_i^{\prime}$ is an indecomposable group code, then $r=s$ and there exists a permutation $\gamma\in S_r$ such that $\Cod{D}_i\simeq \Cod{D}^{\prime}_{\gamma(i)}$ for each $i\in I_r$.
\end{theorem}

\begin{IEEEproof}
If there is a group code $\Cod{C}\subseteq G^{n}$ that can not be written as a sum of indecomposables,  there is one such code of minimal length.  This code cannot be indecomposable and thus there exist $\Cod{D}$, $\Cod{E}$ such that $\Cod{C}\simeq\Cod{D}\oplus \Cod{E}$. Since the lengths of $\Cod{D}$ and $\Cod{E}$ are strictly less than $n$, by assumption $\Cod{D}$ and $\Cod{E}$ are sum of indecomposables. That is, $\Cod{D}\simeq \Cod{A}_1\oplus \cdots \oplus \Cod{A}_r$ and $\Cod{E}\simeq \Cod{B}_1\oplus \cdots \oplus \Cod{B}_s$, where all  $\Cod{A}_i$ and $\Cod{B}_j$ are indecomposable.  Hence, \vspace{-4pt}
\[
\Cod{C}\simeq \Cod{D}\oplus \Cod{E} \simeq \left(\Cod{A}_1\oplus\Cod{A}_2\oplus \cdots \oplus \Cod{A}_r\right) \oplus  \left(\Cod{B}_1\oplus\Cod{B}_2\oplus \cdots \oplus \Cod{B}_s\right), \vspace{-4pt}
\]
a contradiction. For the uniqueness property, assume that
$$\Cod{C}\simeq \Cod{D}_1\oplus\Cod{D}_2\oplus\cdots \oplus \Cod{D}_r\simeq \Cod{D}^{\prime}_1\oplus\Cod{D}^{\prime}_2\oplus\cdots \oplus\Cod{D}^{\prime}_s$$
with all  $\Cod{D}_{i}$ and $\Cod{D}^{\hspace{1pt}\prime}_{i}$ indecomposable group codes, and that $r\leq s$. We do  induction on $r$. If $r=1$, then $\Cod{C}\simeq \Cod{D}_1\simeq \Cod{D}^{\prime}_1\oplus\Cod{D}^{\prime}_2\oplus\cdots \Cod{D}^{\prime}_s$ is indecomposable and we must then have that $s=1$ and $\Cod{D}_1\simeq \Cod{D}^{\prime}_1$. Assume that the result is valid up to $r$ and that there is an isomorphism
$$\varphi:\Cod{D}_1\oplus\Cod{D}_2\oplus\cdots \oplus\Cod{D}_r\oplus\Cod{D}_{r+1}\simeq \Cod{D}^{\prime}_1\oplus\Cod{D}^{\prime}_2\oplus\cdots \oplus \Cod{D}^{\prime}_s$$
If $n$ is the length of $\Cod{C}$, by Proposition \ref{autogroups} there exist $(f_1,\ldots,f_{n})\in \aut(G)^{n}$  and $\overline{\sigma}\in \text{Equ}\left(G^{n}\right)$ such that $\varphi=(f_1,\ldots,f_{n})\circ\overline{\sigma}$.

Let   $I_{_{\Cod{D}_1}}=I_{n_1}\varsubsetneq I_n$ be the set of indexes that label the coordinates of $\Cod{D}_1\subseteq G^{n_1}$ in the sum $\Cod{D}_1\oplus\Cod{D}_2\oplus\cdots \oplus \Cod{D}_{m+1}$. Likewise, let $I_{_{\Cod{D}^{\prime}_i}}\varsubsetneq I_n$ be the set of indexes that label the coordinates of $\Cod{D}^{\prime}_i$ in the sum $\Cod{D}^{\prime}_1\oplus\Cod{D}^{\prime}_2\oplus\cdots\oplus \Cod{D}^{\prime}_s$. Then, $\sigma(I_{_{\Cod{D}_1}})\cap I_{_{\Cod{D}^{\prime}_i}}\neq\emptyset$ for some $i\in I_s$, and we may assume that $\sigma(I_{_{\Cod{D}_1}})\cap I_{_{\Cod{D}^{\prime}_1}}\neq\emptyset$. We claim that  $\sigma(I_{_{\Cod{D}_1}})\subseteq I_{_{\Cod{D}^{\prime}_1}}$. Indeed, if otherwise $\sigma(I_{_{\Cod{D}_1}})\nsubseteq I_{_{\Cod{D}^{\prime}_1}}$, let $J=\sigma(I_{_{\Cod{D}_1}})\cap I_{_{\Cod{D}^{\prime}_1}}$ and $K=\sigma(I_{_{\Cod{D}_1}})- I_{_{\Cod{D}^{\prime}_1}}$. For the identity $\overline{e}$ of  $G^{n_1}$ and the corresponding inclusion $i_{\overline{e}}:\Cod{D}_1\rightarrow \bigoplus _{i=1}^{m}\Cod{D}_i$,  define\vspace{-4pt}
\[
\psi_{_J}=\pi_{_J}\circ \varphi\circ i_{\overline{e}}:\Cod{D}_1\rightarrow \pi_{_J}\left(\varphi\left(i_{\overline{e}}\left(\Cod{D}_1\right)\right)\right)\vspace{-4pt}
\]
\[
\psi_{_K}=\pi_{_K}\circ \varphi\circ i_{\overline{e}}:\Cod{D}_1\rightarrow \pi_{_K}\left(\varphi\left(i_{\overline{e}}\left(\Cod{D}_1\right)\right)\right)
\]
and the group code morphism \vspace{-2pt}
\[
\psi:\Cod{D}_1\rightarrow \pi_{_J}\left(\varphi\left(i_{\overline{e}}\left(\Cod{D}_1\right)\right)\right)\oplus \pi_{_K}\left(\varphi\left(i_{\overline{e}}\left(\Cod{D}_1\right)\right)\right)
\]
given by $\psi\left(x\right)=\left(\psi_{_J}(x) ,\psi_{_K}(x) \right)$ for $x\in G^{n_1}$. Since $\sigma(I_{_{\Cod{D}_1}})=J\cup K$ and  $J\cap K=\emptyset$, then $\pi_{_J}\left(\varphi\left(i_{\overline{e}}\left(\Cod{D}_1\right)\right)\right)\oplus \pi_{_K}\left(\varphi\left(i_{\overline{e}}\left(\Cod{D}_1\right)\right)\right)$ and  $\Cod{D}_1$ have the same length $n_1$ and  $d\left(\psi(x),\psi(y)\right)=d(x,y)$ for all $x,y\in G^{n_1}$. To show that $\psi$ is a group code morphism observe that\vspace{-4pt}
\[
\psi\left(\Cod{D}_1\right)\subseteq \pi_{_J}\left(\varphi\left(i_{\overline{e}}\left(\Cod{D}_1\right)\right)\right)\oplus \pi_{_K}\left(\varphi\left(i_{\overline{e}}\left(\Cod{D}_1\right)\right)\right).
\]
For the other inclusion, if $a\in \pi_{_J}\left(\varphi\left(i_{\overline{e}}\left(\Cod{D}_1\right)\right)\right)$, since $J\subseteq I_{_{\Cod{D}^{\prime}_1}}$, there exists $\alpha\in\Cod{D}_1$ such that $\varphi\left(i_{\overline{e}}\left(\alpha\right)\right)=\left(a_1,\overline{e_1}\right)\in \varphi\left(i_{\overline{e}}\left(\Cod{D}_1\right)\right)$, where $a_1\in\Cod{D}_1^{\prime}$ and $\overline{e_1}$ is the identity of $\bigoplus _{i=2}^{m}\Cod{D}_i$. Therefore $ \pi_{_J}\left(a_1,\overline{e_1}\right)=a$ and  $\pi_{_K}\left(a_1,\overline{e_1}\right)=\overline{e_{_K}}$, where $\overline{e_{_K}}$ is the identity of $\pi_{_K}\left(\varphi\left(i_{\overline{e}}\left(\Cod{D}_1\right)\right)\right)$. If $b\in \pi_{_K}\left(\varphi\left(i_{\overline{e}}\left(\Cod{D}_1\right)\right)\right)$, since $K\cap I_{_{\Cod{D}^{\prime}_1}}=\emptyset$, there exists $\beta\in\Cod{D}_1$ such that $\varphi\left(i_{\overline{e}}\left(\beta\right)\right)=\left(\overline{e_2},b_1\right)\in \varphi\left(i_{\overline{e}}\left(\Cod{D}_1\right)\right)$, where $\overline{e_2}$ is the identity of $\Cod{D}^{\prime}_1$ and $b_1\in\bigoplus _{i=2}^{m}\Cod{D}_i$. Therefore, $ \pi_{_K}\left(\overline{e_2},b_1\right)=b$ and  $\pi_{_K}\left(\overline{e_2},b_1\right)=\overline{e_{_J}}$, where $\overline{e_{_J}}$ is the identity of $\pi_{_J}\left(\varphi\left(i_{\overline{e}}\left(\Cod{D}_1\right)\right)\right)$. Hence, for  $(a,b)\in \pi_{_J}\left(\varphi\left(i_{\overline{e}}\left(\Cod{D}_1\right)\right)\right)\oplus \pi_{_K}\left(\varphi\left(i_{\overline{e}}\left(\Cod{D}_1\right)\right)\right)$, there exist $\alpha,\beta\in\Cod{D}_1$, such that
\begin{align*}
  \psi(\alpha \beta) &= \psi(\alpha )\psi(\beta)= \left(\psi_{_J}(\alpha) ,\psi_{_K}(\alpha) \right) \left(\psi_{_J}(\beta) ,\psi_{_K}(\beta) \right) \\
   &=\left(a,\overline{e_{_K}}\right) \left(\overline{e_{_J}}, b\right) = (a,b)
    \end{align*}
Thus, $\psi$ is a group code isomorphism and hence $\Cod{D}_1$ is a decomposable code, a contradiction. It follows that
$\sigma(I_{_{\Cod{D}_1}})\subseteq I_{_{\Cod{D}^{\prime}_1}}$. A similar argument, for $\varphi^{-1}=(f_{_{{\sigma}^{\hspace{1pt}-1}}})^{-1} \circ \overline{\sigma}^{\hspace{1pt}-1}$ shows that $\sigma^{-1}( I_{_{\Cod{D}^{\prime}_1}}) \subseteq I_{_{\Cod{D}_1}}$, and hence $\sigma(I_{_{\Cod{D}_1}})= I_{_{\Cod{D}^{\prime}_1}}$. From Proposition \ref{lemcanisomor}, $\Cod{D}_1\simeq \Cod{D}^{\prime}_1$ and $\Cod{D}_2\oplus\cdots \oplus \Cod{D}_{m+1} \simeq \Cod{D}^{\prime}_2\oplus\cdots\oplus \Cod{D}^{\prime}_s$. By the induction hypothesis the result follows.
\end{IEEEproof}

For each $j\in I_m$, consider a group code $\Cod{D}_j \subseteq G^{n_j}$ and the direct sum group code  $\bigoplus _{j=1}^{m}\Cod{D}_j\subseteq G^{n}$, where $n=\sum _{j=1}^{m}n_j$. Let $\overline{e_j}$ be the identity of  $G^{n_j}$ and  $i_{\overline{e_j}}$ the corresponding inclusion of $\Cod{D}_j$ in $\bigoplus _{j=1}^{m}\Cod{D}_j$. We denote its image by $i_{\overline{e_j}}(\Cod{D}_j)=\widetilde{\Cod{D}_j}$.

\begin{corollary}\label{resdemor_0}
Let  $\bigoplus_{j=1}^{m}\Cod{D}_j\subseteq G^{n}$ be a direct sum of indecomposable group codes and $\varphi=f\circ \overline{\sigma}\in \aut_{_{GC}}\big(\bigoplus_{j=1}^{m}\Cod{D}_j\big)$, with $f\in (\aut(G))^{n}$ and $\overline{\sigma}\in \text{Equ}(G^{n})$. If  $I_{\Cod{D}_j}\subseteq I_n$ is the set of indexes that label the elements of  $\Cod{D}_j$ in the sum $\bigoplus_{j=1}^{m} \Cod{D}_j$, then $\sigma(I_{_{\Cod{D}_j}})\cap I_{_{\Cod{D}_k}}\neq \emptyset$ if and only if $\varphi\left(\widetilde{\Cod{D}_j}\right)=\widetilde{\Cod{D}_k}$.
\end{corollary}

If  $\Cod{E}\subseteq G^{n}$ is an indecomposable group code and $\alpha\in \Z^{+}$, we use the notation $\Cod{E}^\alpha=\Cod{E}_1\oplus\cdots \oplus \Cod{E}_\alpha$, where $\Cod{E}_j=\Cod{E}$ for each $j\in I_\alpha$. Thus, if $\bigoplus_{i=1}^{m}\Cod{D}_i\subseteq G^{n}$ is a direct sum of indecomposable group codes, joining together the isomorphic group codes and reindexing we may write\vspace{-8pt}
\[
\bigoplus _{i=1}^{m}\Cod{D}_i\simeq \bigoplus _{j=1}^{k}\Cod{E}_j^{\alpha_{j}},\vspace{-4pt}
\]
where $\Cod{E}_j\simeq\Cod{D}_{i_j}$ for some $i_j\in I_m$ and  $\Cod{E}_s \nsimeq \Cod{E}_t$ if $s\neq t$.

\begin{corollary}\label{resdemor}
Let  $\bigoplus_{j=1}^{k}\Cod{D}_j^{\alpha_{j}}$ be  a direct sum of indecomposable group codes, with $\Cod{D}_s\nsimeq \Cod{D}_t$ for  $s\neq t$. If  $\varphi\in \aut_{_{GC}}\left(\bigoplus_{j=1}^{k}\Cod{D}_j^{\alpha_{j}}\right)$, then  $\varphi\Big(\widetilde{\Cod{D}_j^{\alpha_{j}}}\Big)=\widetilde{\Cod{D}_j^{\alpha_{j}}}$.
\end{corollary}

\begin{proposition}\label{isoauto_1}
If  $\bigoplus_{j=1}^{k}\Cod{D}_j^{\alpha_{j}}$ is a direct sum of indecomposable group codes, where $\Cod{D}_s\nsimeq \Cod{D}_t$ for $s\neq t$, then $\aut_{GC}\left(\bigoplus_{j=1}^{k}\Cod{D}_j^{\alpha_{j}}\right)$ is a group isomorphic to $\prod_{j=1}^{k}  \aut_{GC}\left( \Cod{D}_j^{\alpha_{j}}\right)$.
\end{proposition}

\begin{IEEEproof}
For each $j\in I_k$ let $I_{_{\Cod{D}_j}}$ be the set of indexes that label the elements of $\Cod{D}_j^{\alpha_j}$ in the sum $\bigoplus_{j=1}^{k}\Cod{D}_j^{\alpha_{j}}$. Let $i_{\overline{e_j}}$ be the inclusion of $\Cod{D}_j^{\alpha_j}$ in the sum $\bigoplus_{j=1}^{k}\Cod{D}_j^{\alpha_{j}}$. The following diagram commutes\vspace{-3pt}
\[
\xymatrix@C=1.5cm{ \Cod{D}_j^{\alpha_{j}} \ar[r]_{i_{\overline{e_j}}} \ar@/^2.5pc/[rrr]^{\varphi_j} &  \bigoplus\limits_{j=1}^{k}\Cod{D}_j^{\alpha_{j}}  \ar[r]_{\varphi} & \bigoplus\limits_{j=1}^{k}\Cod{D}_j^{\alpha_{j}} \ar[r]_{\pi_{_{I_{_{\Cod{D}_j}}}}} &  \Cod{D}_j^{\alpha_{j}} }\vspace{-3pt}
\]
By Corollary \ref{resdemor}, $\varphi_j= \pi_{_{\Cod{D}_j}} \circ \varphi \circ i_{\overline{e_j}}$ is an automorphism of the group code $\Cod{D}_j^{\alpha_{j}}$, for each $j\in I_k$. Since $i_{\overline{e_j}} \circ \pi_{_{I_{_{\Cod{D}_j}}}}=\id_{_{\widetilde{\Cod{D}_j^{\alpha_{j}}}}}: \widetilde{\Cod{D}_j^{\alpha_{j}}}\rightarrow \widetilde{\Cod{D}_j^{\alpha_{j}}}$, if $\varphi,\psi\in \aut_{GC}\left(\bigoplus_{j=1}^{k}\Cod{D}_j^{\alpha_{j}}\right)$ , then \vspace{-8pt}
\begin{align*}
(\psi \circ \varphi)_j&= \pi_{_{I_{_{\Cod{D}_j}}}} \circ \psi \circ \varphi \circ i_{\overline{e_j}}=\pi_{_{I_{_{\Cod{D}_j}}}} \circ \psi \circ i_{\overline{e_j}} \circ \pi_{_{I_{_{\Cod{D}_j}}}} \circ \varphi \circ i_{\overline{e_j}}  \\
                         & =\psi_j \circ \varphi_j.
    \end{align*}
This shows that the map
$$\chi: \aut_{_{GC}}\Big(\bigoplus_{j=1}^{k}\Cod{D}_j^{\alpha_{j}}\Big) \rightarrow \prod_{j=1}^{k}  \aut_{_{GC}}\big(\Cod{D}_j^{\alpha_{j}}\big)$$
given by $\varphi \mapsto (\varphi_1,\ldots,\varphi_k)$ is a homomorphism of groups, which clearly is an isomorphism.
\end{IEEEproof}

\begin{proposition}\label{isoauto_2}
Let $\Cod{D}\subseteq G^{m}$ be an indecomposable group code and $\alpha\in \Z^{+}$. Then,  $\aut_{_{GC}}\left(\Cod{D}^{\hspace{1pt}\alpha}\right)$ is a group isomorphic to $ \aut_{_{GC}}\left( \Cod{D}\right) ^{\alpha}\rtimes S_{\alpha}$.
\end{proposition}

\begin{IEEEproof}
$\Cod{D}^{\alpha}=\bigoplus_{i=1}^{\alpha}\Cod{D}_i$, where $\Cod{D}_i=\Cod{D}$ for each $i\in I_{\alpha}$. By Corollary \ref{resdemor_0} for each $j\in I_\alpha$ there exists a unique $k_j\in I_\alpha$ such that $\varphi(\widetilde{\Cod{D}_j})=\widetilde{\Cod{D}_{k_j}}$ Let $\gamma:I_\alpha\rightarrow I_\alpha$ be the permutation given by $\gamma(j)=k_j$ and let
$\overline{\gamma}^{\hspace{1pt}-1}\in \text{Equ}\left(G^{n}\right)$ be defined by $\overline{\gamma}^{\hspace{1pt}-1}\left(a_1,\ldots,a_\alpha\right)=\left(a_{\gamma^{\hspace{1pt -1}}(1)},\ldots,a_{\gamma^{\hspace{1pt}-1}}(\alpha)\right)$ for $(a_1,\ldots,a_\alpha)\in \left(G^{m}\right)^{\hspace{1pt}\alpha}$, where $a_j\in G^{m}$ for $j\in I_{\alpha}$. Note that $\overline{\gamma}\left(\widetilde{\Cod{D}_j}\right)=\widetilde{\Cod{D}_{\gamma^{\hspace{1pt}-1}(j)}}$. For each $j\in I_\alpha$, let $i_j:\Cod{D}\rightarrow \Cod{D}^{\hspace{1pt}\alpha}$ be the $j$-th inclusion of $\Cod{D}$ in $\Cod{D}^{\hspace{1pt}\alpha}$, and let $I_{_{\Cod{D}_j}}$ be the set of indexes of the $j$-th summand of $\Cod{D}^{\hspace{1pt}\alpha}$. We then have a commutative diagram\vspace{-4pt}
\[
\xymatrix{ \Cod{D} \ar[r]_{i_j} \ar@/^1.5pc/[rrr]^{\varphi_j} & \Cod{D}^\alpha  \ar[r]_{\varphi \circ \overline{\gamma}^{\vspace{1pt}-1}}  & \Cod{D}^\alpha \ar[r]_{\pi_{_{I_{_{\Cod{D}_j}}}}} &  \Cod{D}}
\]
where $\varphi_j= \pi_{_{I_{j}}} \circ \varphi \circ \overline{\gamma} \circ i_j$ is an automorphism of group codes of $\Cod{D}$. Clearly, $ \varphi \circ \overline{\gamma}^{\hspace{1pt}-1} = \bigoplus_{j=1}^{k}\varphi_j$.

If  $H=\big\{\bigoplus_{j=1}^{k}\varphi_j\in \aut_{_{GC}}(\Cod{D}^\alpha): (\varphi_1,\ldots,\varphi_\alpha)\in (\aut_{_{GC}}(\Cod{D}))^{\alpha}\big\}$ and   $N=\{\overline{\gamma}\in \aut_{_{GC}}(\Cod{D}^\alpha)\mid \gamma\in I_\alpha \}$, then $H$ and $N$ are subgroups of $\aut_{_{GC}}(\Cod{D}^\alpha)$. A direct computation shows that
\[
\overline{\gamma}^{\vspace{1pt}-1} \circ\bigoplus\limits_{j=1}^{k}\varphi_j \circ \overline{\gamma}= \bigoplus\limits_{j=1}^{k}\varphi_{\gamma^{\vspace{1pt}-1}(j)}
\]
for all elements $\bigoplus_{j=1}^{k}\varphi_j\in H$ and $\overline{\gamma}\in N$. Therefore, $H$ is a normal subgroup of $\aut_{_{GC}}(\Cod{D})$ and since $H \cap N = Id_{\Cod{D}^{\vspace{1pt}\alpha}}$, then $\aut_{_{GC}}\Cod(D)^{\alpha}=H\rtimes N$. Since $H\simeq(\aut_{_{GC}}\Cod(D))^{\alpha}$ and $N\simeq S_\alpha$, the result follows.
\end{IEEEproof}

Propositions \ref{isoauto_1} and  \ref{isoauto_2} give:

\begin{theorem}
If $\bigoplus_{j=1}^{k}\Cod{D}_j^{\alpha_{j}}$ is a direct sum of indecomposable group codes, where $\Cod{D}_s\nsimeq \Cod{D}_t$ for $s\neq t$, then $\aut_{_{GC}}\big(\bigoplus_{j=1}^{k}\Cod{D}_j^{\alpha_{j}}\big)$ is a group isomorphic to $\prod_{j=1}^{k} \big((\aut{_{_{GC}}}\left( \Cod{D}\right)) ^{\alpha_j}\rtimes S_{\alpha_j}\big)$.
\end{theorem}

The following result gives examples of indecomposable group codes. We say that $\Cod{C}\subseteq G^{n}$ is a {\it constant weight group code} if there exists an integer $0<r\leq n$, such that $\Cod{C}-\{\overline{e}_n\}\subseteq B_{\overline{e}_n}^{r}$,  where $\overline{e}_n$ is the identity of $G^n$.  If $x\in G^{n}$, the  \textit{weight} of $x$ in $G^{n}$ is $w(x)=d(x,\overline{e}_n)$. Then, $\Cod{C}\subseteq G^{n}$
is a constant weight code of weight  $r$ if and only if $w(x)=r$, for all $x\in \Cod{C}-\{\overline{e}_n\}$.

\begin{proposition}
Every non degenerated constant weight code is indecomposable.
\end{proposition}

\begin{IEEEproof}\label{weconsines}
Assume that there exist a group code $\Cod{C}\subseteq G ^{n}$ of constant weight $r$ and $\Cod{D}\subseteq G^{k}$ y $\Cod{E}\subseteq G^{l}$ group codes such that $\Cod{C}\simeq\Cod{D}\oplus \Cod{E}$. Since $\Cod{C}$ is non degenerated, then $\left|\Cod{D}\right|\geq 2$ and  $\left|\Cod{E}\right|\geq 2$. Hence, there exist $a_1,a_2\in\Cod{D}$ and $b_1,b_2\in\Cod{E}$, with $a_1\neq a_2$ and $b_1\neq b_2$. Since
\begin{align*}
r&=w((a_1a_2^{-1},b_1b_2^{-1}))=d_{_{G^{n}}}\left((a_1,b_1),(a_2,b_2)\right)\\
&=d_{_{G^{m}}}\left(a_1,a_2\right)+d_{_{G^{l}}}\left(b_1,b_2\right) \vspace{-2pt}
\end{align*}
and
$$r=w((a_1a_2^{-1},b_1b_1^{-1}))=d_{_{G^{n}}}\left((a_1,b_1)(a_2,b_1)\right)=d_{_{G^{m}}}\left(a_1,a_2\right)$$
then $d_{_{G^{l}}}\left(b_1,b_2\right)=0$ and so $b_1=b_2$, a contradiction.
\end{IEEEproof}

\section{The Structure of Cyclic Group Codes}
For any finite alphabet $A$, a {\it cyclic code} is a code $\Cod{C}\subseteq A^{n}$  such that for all   $c\in \Cod{C}$ and for the $n$-cycle $\delta=(1,\ldots,n)\in S_n$ for the equivalence $\overline{\delta}$ we have that $\overline{\delta}(c)\in \Cod{C}$. A {\it cyclic group code} is a cyclic code $\Cod{C}\subseteq G^{n}$ which is also a subgroup of $G^n$.

\begin{theorem}\label{estrucicli}
Let  $\Cod{C}\subseteq G^{n}$ be a decomposable cyclic group code, say
$\Cod{C}\simeq \bigoplus_{j=1}^{m}\Cod{D}_j$, with $\Cod{D}_j$ indecomposable group codes for each $j\in I_m$. Then, $\Cod{D}_j\simeq\Cod{D}_1$ for each $j\in I_m$.
\end{theorem}

\begin{IEEEproof}
Let  $\varphi:\Cod{C}\rightarrow \bigoplus_{j=1}^{m}\Cod{D}_j$ be an isomorphism of group codes. Let  $f\in (\aut(G))^n$ and $\overline{\sigma}\in\text{Equ}\left(G^{n}\right)$ such that $\varphi=f\circ \overline{\sigma}$. For the cycle $\delta=(12\ldots n)\in S_n$, since $\Cod{C}\subseteq G^{n}$ is  a cyclic group code, then $\overline{\delta}^{\hspace{2pt}t}\in\aut_{_{GC}}(C)$ for all $t\in \N$. Hence, $\varphi\circ \overline{\delta}^{\hspace{2pt}t}\circ\varphi^{-1}\in\aut_{_{GC}}\big(\bigoplus_{j=1}^{m}\Cod{D}_j\big)$ for all $t\in \N$, and
\begin{align*}
\varphi\circ \overline{\delta}^{\hspace{2pt}t}\circ\varphi^{-1}&= f\circ \overline{\sigma} \circ \overline{\delta}^{\hspace{2pt}t}\circ \big(f_{_{{\sigma}^{\hspace{1pt}-1}}}\big)^{-1} \circ \overline{\sigma}^{\hspace{1pt}-1}  \vspace{2pt} \\
   &=f\circ  \big(\big(f_{_{{\sigma}^{\hspace{1pt}-1}}}\big)^{-1}\big)_{_{\sigma \circ \delta^{\hspace{1pt}t}}}  \circ  \overline{\sigma} \circ \overline{\delta}^{\hspace{2pt}t}\circ\overline{\sigma}^{\hspace{1pt}-1}
\end{align*}
For each $j\in I_m$ let $I_{_{\Cod{D}_j}}$ be the set of indexes that label the elements of  $\Cod{D}_j$ in the sum $\bigoplus_{j=1}^{m}\Cod{D}_j$. If $k_j\in I_{_{\Cod{D}_j}}$, there exists $t_j\in \N$ such that $(\delta^{\hspace{1pt}t_j}\circ \sigma^{-1})(1)=\sigma^{-1}(k_j)$ or equivalently  $(\sigma\circ \delta^{\hspace{1pt}t_j}\circ \sigma^{-1})(1)=k_j$. That is, $(\sigma\circ \delta^{\hspace{1pt}t_j}\circ \sigma^{-1})(I_{_{\Cod{D}_1}})\cap I_{_{\Cod{D}_j}}\neq\emptyset$, and thus, by Corollary \ref{resdemor_0}, $\Cod{D}_j\simeq\Cod{D}_1$ for each $j\in I_m$.
\end{IEEEproof}

\begin{corollary}\label{inescodcy}
Let  $\Cod{C}\subseteq G^{n}$ be a cyclic group code and write its order as $\left|\Cod{C}\right|=p_1^{\xi_1}\cdots p_s^{\xi_s}$ with $p_i$ prime integers. Let $\xi=\gcd(\xi_1,\ldots,\xi_n)$ be its greatest common factor. If  $\gcd(\xi,n)=1$, then $\Cod{C}$ is an indecomposable group code.
\end{corollary}

\begin{IEEEproof}
If  $\Cod{C}\subseteq G^{n}$ is a decomposable cyclic group code, by Proposition \ref{estrucicli}, $\Cod{C}\simeq\Cod{D}^{\alpha}$, where  $\Cod{D}\subseteq G^{m}$ is an indecomposable group code and $\alpha\geq 2$. Since the lengths of $\Cod{C}$ and $\Cod{D}^{\alpha}$ are the same, then $m\alpha=n$.  Writing $\left|\Cod{D}\right|=p_1^{\zeta_1}\cdots p_s^{\zeta_s}$, then $\zeta_i \alpha=\xi_i$, and thus $\alpha\geq 2$ divides $\xi$. Since by hypothesis $\gcd(\xi,n)=1$, then $\alpha=1$, a contradiction.
\end{IEEEproof}

\begin{example}
The converse of Corollary \ref{inescodcy} is false. Indeed, if  $G$ is a finite group such that $\left|G^{n}\right|=p_1^{\xi_1}\cdots p_s^{\xi_s}$ with $\xi=\gcd(\xi_1,\ldots,\xi_n)$, then $G^{n}$ is a decomposable group code for all $n\geq 2$, in particular  for $n$ such that $\gcd(\xi,n)=1$.
\end{example}

\begin{proposition}\label{creacodcicli}
If $\Cod{D}\subseteq G^{m}$ is a cyclic group code, then for any nonnegative integer $\ell$, $\Cod{D}^{\ell}$ is isomorphic to a cyclic group code.
\end{proposition}

\begin{IEEEproof}
Put $n=\ell m$. Every $t\in I_n$ can be written in a unique way as $t=sm+r$, with $0\leq s\leq \ell -1$ and $1\leq r\leq m$. Define $\sigma\in S_n$ by $\sigma(t)=\sigma(sm+r)=(r-1)\ell+(s+1)$. Consider the cycle $\delta=(1\cdots n)\in S_n$ and \vspace{-5pt}
{\fontsize{9}{12}\[
\left((a_{11},a_{12},\ldots,a_{1m}),(a_{21},a_{22},\ldots,a_{2m}),\ldots,(a_{\ell 1},a_{\ell 2},\ldots,a_{\ell m})\right)\in \Cod{D}^{\ell}.
\]}\vspace{-6pt}
Then,
\begin{multline*}
\!\!\! \overline{\delta}\left(\overline{\sigma}\left((a_{11},\ldots,a_{1m}),(a_{21},\ldots,a_{2m}),\ldots,(a_{\ell 1},\ldots ,a_{\ell m})\right)\right) \\
= \overline{\delta}\left(a_{11},\ldots,a_{\ell 1},a_{12},\ldots,a_{\ell 2},\ldots,a_{1 m},\ldots,a_{\ell-1 m},a_{\ell m}\right)  \\
\!\!\!\!\!\!\!\!\!\!\!\!\!\!\!=(\underset{\ell-places }{\underbrace{a_{\ell m},a_{11},\ldots,a_{(\ell-1) 1}}},\underset{\ell-places }{\underbrace{a_{\ell 1},a_{12},\ldots,a_{(\ell-1) 2}}},\ldots, \\
   \qquad\qquad\qquad\qquad\qquad\qquad\underset{\ell-places }{\underbrace{a_{\ell ( m-1)},a_{1 m},\ldots,a_{\ell-1 m}}} )\\
\!\!\!\!\!\!\!\!\! = \overline{\sigma}((a_{\ell m},a_{\ell 1},\ldots,a_{\ell (m-1)}),(a_{11},a_{12},\ldots,a_{1m}),\ldots,\\
 \qquad\qquad\qquad\qquad\qquad\qquad (a_{(\ell-1) 1},a_{(\ell-1) 2},\ldots,a_{(\ell-1) m})).
\end{multline*}
Therefore, $\Cod{C}=\overline{\sigma}(\Cod{D}^{\ell})$ is a cyclic group code.
\end{IEEEproof}

\begin{example}
For $G=\Z/2$, the group of integers modulo $2$, consider the group code
$$\Cod{D}=\{(0,0,0),(\overline{1},\overline{1},0),(0,\overline{1},\overline{1}),(\overline{1},0,\overline{1})\}\subseteq (\Z/2)^{3}.$$
By Proposition \ref{creacodcicli}, $\Cod{D}^{2}\subseteq (\Z/2)^6$ is isomorphic to a cyclic group code  $\Cod{C}$ by means of the equivalence $\overline{\sigma}$ whose permutation $\sigma\in S_6$ is given by $\sigma(1)=1$, $\sigma(2)=3$, $\sigma(3)=5$, $\sigma(4)=2$, $\sigma(5)=4$, $\sigma(6)=6$.\vspace{3pt}

\begin{center}
\begin{tabular}{c @{\ \ } c @{\ \ } c}
  $\Cod{D}^{2}$ & & $\Cod{C}$ \\
  \hline
  $(\mathbf{0},\mathbf{0},\mathbf{0},0,0,0)$ & $\mapsto$ & $(\mathbf{0},0,\mathbf{0},0,\mathbf{0},0)$ \\
  $(\mathbf{0},\mathbf{0},\mathbf{0},1,1,0)$ & $\mapsto$ & $(\mathbf{0},1,\mathbf{0},1,\mathbf{0},0)$ \\
  $(\mathbf{0},\mathbf{0},\mathbf{0},0,1,1)$ & $\mapsto$ & $(\mathbf{0},0,\mathbf{0},1,\mathbf{0},1)$ \\
  $(\mathbf{0},\mathbf{0},\mathbf{0},1,0,1)$ & $\mapsto$ & $(\mathbf{0},1,\mathbf{0},0,\mathbf{0},1)$ \\
  $(\mathbf{1},\mathbf{1},\mathbf{0},0,0,0)$ & $\mapsto$ & $(\mathbf{1},0,\mathbf{1},0,\mathbf{0},0)$ \\
  $(\mathbf{1},\mathbf{1},\mathbf{0},1,1,0)$ & $\mapsto$ & $(\mathbf{1},1,\mathbf{1},1,\mathbf{0},0)$ \\
  $(\mathbf{1},\mathbf{1},\mathbf{0},0,1,1)$ & $\mapsto$ & $(\mathbf{1},0,\mathbf{1},1,\mathbf{0},1)$ \\
  $(\mathbf{1},\mathbf{1},\mathbf{0},1,0,1)$ & $\mapsto$ & $(\mathbf{1},1,\mathbf{1},0,\mathbf{0},1)$ \\
  $(\mathbf{0},\mathbf{1},\mathbf{1},0,0,0)$ & $\mapsto$ & $(\mathbf{0},0,\mathbf{1},0,\mathbf{1},0)$ \\
  $(\mathbf{0},\mathbf{1},\mathbf{1},1,1,0)$ & $\mapsto$ & $(\mathbf{0},1,\mathbf{1},1,\mathbf{1},0)$ \\
  $(\mathbf{0},\mathbf{1},\mathbf{1},0,1,1)$ & $\mapsto$ & $(\mathbf{0},0,\mathbf{1},1,\mathbf{1},1)$ \\
  $(\mathbf{0},\mathbf{1},\mathbf{1},1,0,1)$ & $\mapsto$ & $(\mathbf{0},1,\mathbf{1},0,\mathbf{1},1)$ \\
  $(\mathbf{1},\mathbf{0},\mathbf{1},0,0,0)$ & $\mapsto$ & $(\mathbf{1},0,\mathbf{0},0,\mathbf{1},0)$ \\
  $(\mathbf{1},\mathbf{0},\mathbf{1},1,1,0)$ & $\mapsto$ & $(\mathbf{1},1,\mathbf{0},1,\mathbf{1},0)$ \\
  $(\mathbf{1},\mathbf{0},\mathbf{1},0,1,1)$ & $\mapsto$ & $(\mathbf{1},0,\mathbf{0},1,\mathbf{1},1)$ \\
  $(\mathbf{1},\mathbf{0},\mathbf{1},1,0,1)$ & $\mapsto$ & $(\mathbf{1},1,\mathbf{0},0,\mathbf{1},1)$ \\
\end{tabular}
\end{center}
\vspace{4pt}
\end{example}

\begin{proposition}
Every decomposable cyclic group code is isomorphic to a direct sum of indecomposable cyclic group codes.
\end{proposition}

\begin{IEEEproof}
Let $\Cod{C}\subseteq G^{n}$ be a decomposable cyclic group code. By Theorem \ref{estrucicli}, there exists an indecomposable group code $\Cod{D}\subseteq G^{m}$  such that $\Cod{C}\simeq \Cod{D}^{\alpha}$. I.e., there exists an isomorphism of group codes $\varphi:\Cod{D}^{\alpha}\rightarrow \Cod{C}$, with $\varphi=f\circ\overline{\sigma}$ and where $f\in (\aut(G ))^{n}$ and  $\overline{\sigma}\in\text{Equ} (G^{n} )$. Recall that $\Cod{D}^{\alpha}=\Cod{D}_1\oplus\cdots\oplus \Cod{D}_\alpha$ with $\Cod{D}_i=\Cod{D}$ for each $i\in I_\alpha$. Let $I_{_{\Cod{D}_i}}$ be the set of indexes that label the coordinates of  $\Cod{D}_i$ in $\Cod{D}^{\alpha}$. For each  $i\in I_\alpha$, let $J_{_{\Cod{D}_i}}=\sigma(I_{_{\Cod{D}_i}})$. If $\delta=(1\cdots n)\in S_n$ and $K_i=\left\{\overline{\delta}^{\hspace{1pt} t}\mid \delta^{ t}(J_{_{\Cod{D}_i}})= J_{_{\Cod{D}_i}} \right\}$. Since $\varphi^{-1}\circ \overline{\delta}^{\hspace{2pt}s}\circ\varphi\in\aut_{_{GC}} (\Cod{D}^{\alpha} )$ for all $s\in\N$, and $\varphi^{-1}\circ \overline{\delta}^{\hspace{2pt}t}\circ\varphi=  (f_{_{{\sigma}^{\hspace{1pt}-1}}})^{-1}\circ  f_{_{\sigma^{-1} \circ \delta^{\hspace{2pt}t}}}  \circ  \overline{\sigma}^{\hspace{1pt}-1} \circ \overline{\delta}^{\hspace{2pt}t}\circ\overline{\sigma}$, we have that $\overline{\delta}^{\hspace{2pt}t}\in K_i$ if and only if $ (\sigma^{-1} \circ\delta^{\hspace{1pt}t} \circ\sigma )  (I_{_{\Cod{D}_i}} )=I_{_{\Cod{D}_i}}$.  $K_i$ acts naturally on $I_{_{\Cod{D}_i}}$ by $\overline{\delta}^{\hspace{1pt} t}\in K_i$ and $j\in I_{_{\Cod{D}_i}}$, $\overline{\delta}^{\hspace{1pt} t}\cdot j:=(\sigma^{-1} \circ\delta^{\hspace{1pt}t} \circ\sigma)(j)$. Since $\Cod{C}$ is a cyclic group code, if $j\in I_{_{\Cod{D}_i}}$, its orbit is $\text{Orb}(j)=I_{_{\Cod{D}_i}}$ and its stabilizer is $\text{Stab}(j)=\id_{G^{n}}$. From the orbit-stabilizer theorem it follows that  $ |K_i |=m$. Moreover, since $H$ is a cyclic group, then $K_i=\langle\overline{\delta}^{\hspace{2pt}t_0}\rangle$, where $t_0$ divides  $n$. Therefore, the order of this element is $o(\overline{\delta}^{\hspace{2pt}t_0})= |K_i |=m$. It then follows that $t_0=\alpha$, and thus $K_i=\langle\overline{\delta}^{\hspace{2pt}\alpha}\rangle$. Hence, if  $j_i=\min J_{_{\Cod{D}_i}}$, then $J_{_{\Cod{D}_i}}=\{j_i,\delta^{\hspace{1pt}\alpha} (j_i ),\delta^{\hspace{1pt}2\alpha} (j_i ),\ldots,\delta^{\hspace{1pt}(m-1)\alpha} (j_i )\}$. The composition
 $\pi_{_{J_{_{\Cod{D}_i}}}} \circ\varphi \circ i_{_{\Cod{D}_i}}:\Cod{D}_i\rightarrow \pi_{_{J_{_{\Cod{D}_i}}}}(\Cod{C})$ (where $i_{_{\Cod{D}_i}}$ is the inclusion of $\Cod{D}_i$ in  $\Cod{D}^{\alpha}$) is an isomorphism of group codes. Therefore $\Cod{C}\simeq \pi_{_{J_{_{\Cod{D}_1}}}}(\Cod{C})\oplus\cdots\oplus \pi_{_{J_{_{\Cod{D}_\alpha}}}}(\Cod{C})$. It remains to show that each summand $\pi_{_{J_{_{\Cod{D}_i}}}}(\Cod{C})$  is a cyclic group code.
To do this, observe that for every $(a_{j_i},a_{\delta^{\hspace{1pt}\alpha}(j_i)},a_{\delta^{\hspace{1pt}2\alpha}(j_i)},\ldots,a_{\delta^{\hspace{1pt}(m-1)\alpha}(j_i)})\in \pi_{_{J_{_{\Cod{D}_i}}}}(\Cod{C})$ there exists $(\ldots,a_{j_i},\ldots,a_{\delta^{\hspace{1pt}\alpha}(j_i)},\ldots,a_{\delta^{\hspace{1pt}2\alpha}(j_i)},\ldots,a_{\delta^{\hspace{1pt}(m-1)\alpha}(j_i)},\ldots)\in\Cod{C}$ such that \vspace{4pt}
\begin{multline*}
\pi_{_{J_{_{\Cod{D}_i}}}}(\ldots,a_{j_i},\ldots,a_{\delta^{\hspace{1pt}\alpha}(j_i)},\ldots,a_{\delta^{\hspace{1pt}2\alpha}(j_i)},\ldots,a_{\delta^{\hspace{1pt}(m-1)\alpha}(j_i)},\ldots)\\
= (a_{j_i},a_{\delta^{\hspace{1pt}\alpha}(j_i)},a_{\delta^{\hspace{1pt}2\alpha}(j_i)},\ldots,a_{\delta^{\hspace{1pt}(m-1)\alpha}(j_i)}).
\end{multline*}
And since
\begin{multline*}
\!\!\!\!\!\pi_{_{J_{_{\Cod{D}_i}}}}( \overline{\delta}^{\hspace{2pt}(m-1)\alpha} (\ldots,a_{j_i},\ldots,a_{\delta^{\hspace{1pt}\alpha}(j_i)},\ldots,a_{\delta^{\hspace{1pt}2\alpha}(j_i)},\ldots,a_{\delta^{\hspace{1pt}(m-1)\alpha}(j_i)},\ldots))\\
= \pi_{_{J_{_{\Cod{D}_i}}}}(\ldots,a_{\delta^{\hspace{1pt}(m-1)\alpha}(j_i)},\ldots,a_{j_i},\ldots,a_{\delta^{\hspace{1pt}\alpha}(j_i)},\ldots,a_{\delta^{\hspace{1pt}(m-2)\alpha}(j_i)},\ldots)\\
= (a_{\delta^{\hspace{1pt}(m-1)\alpha}(j_i)},a_{j_i},a_{\delta^{\hspace{1pt}\alpha}(j_i)},\ldots,
a_{\delta^{\hspace{1pt}(m-2)\alpha}(j_i)})
\end{multline*}
it follows that $\pi_{_{J_{_{\Cod{D}_i}}}} (\Cod{C} )$ is a cyclic code for every $i\in I_{\alpha}$. Since the lengt of $\pi_{_{J_{_{\Cod{D}_i}}}} (\Cod{C} )$ is $m$ for every $i\in I_{\alpha}$, by Theorem \ref{estrucdecoddege} $\pi_{_{J_{_{\Cod{D}_i}}}} (\Cod{C} )$ is an indecomposable code for every $i\in I_{\alpha}$.
\end{IEEEproof}

\begin{example}
Let  $n,m\in \Z^{+}$ and for each $i\in I_m$ let $G_i$ be a finite group and $G=\prod_{i=1}^{m}G_i$. If $\Cod{C}_i\subseteq G_i^{n}$ is a cyclic group code for each $i\in I_m$, we define the \textit{join} of the family of group codes $ \{\Cod{C}_i\}_{i=1}^{m}$ as the group code $\coprod_{i=1}^m\Cod{C}_i=\{ ((h_{11},h_{21},\ldots,h_{m1}),\ldots,(h_{1n},h_{2n},\ldots,h_{mn}))\in G^{n} :   (h_{i1},h_{i2},\ldots,h_{in})\in \Cod{C}_i\}$,
which clearly is a cyclic group code.
\end{example}

\section{Conclusions}

With the definition of morphism of codes that we introduced for arbitrary group codes, the concept of isomorphism of codes coincides with the classical one for linear codes over Frobenius rings, in particular for linear codes over finite fields. All classification results are generalized in the new context with streamlined proofs.

\end{document}